\documentclass[]{wispaper}
\usepackage{helvet}

\usepackage{amsmath} 
\usepackage{natbib}
\usepackage{graphicx}
\usepackage{subcaption} 

\usepackage[toc,page,header]{appendix}
\usepackage[utf8]{inputenc} 
\usepackage[T1]{fontenc}    
\usepackage{hyperref}       
\usepackage{url}            
\usepackage{booktabs}       
\usepackage{lmodern}        
\usepackage{amsfonts}       
\usepackage{nicefrac}       
\usepackage{microtype}      
\usepackage{wrapfig}

\usepackage{amssymb}  
\usepackage{fontawesome}  
\usepackage{url}  

\usepackage{titletoc}

\usepackage{tikz}  
\usepackage{comment}  
\usepackage{tabularx}  
\usepackage{booktabs}  

\usepackage{minitoc}

\usepackage{booktabs}
\usepackage{array}
\usepackage{etoolbox}

\definecolor{lightblue}{RGB}{200, 230, 255}  
\definecolor{headerblue}{RGB}{150, 200, 255} 

\usepackage{pgfplots}
\usepackage[utf8]{inputenc} 
\usepackage[T1]{fontenc}    
\usepackage{hyperref}       
\usepackage{url}            
\usepackage{booktabs}       
\usepackage{amsfonts}       
\usepackage{nicefrac}       
\usepackage{microtype}      
\usepackage{xcolor}         
\usepackage{graphicx}
\usepackage{float}
\usepackage{comment}
\usepackage{multirow} 
\usepackage{amsmath} 
\usepackage{makecell} 
\usepackage{siunitx}  
\usepackage{tikz}
\usepackage{pgf-pie} 
\usepackage{subcaption}
\usepackage{wrapfig}
\usepackage[export]{adjustbox}

\usepackage{ragged2e}      
\usepackage{tabularx}       
\usepackage{array}          
\usepackage{caption}        
\usepackage{enumitem}
\usepackage{pifont}
\usepackage[hang,flushmargin]{footmisc} 

\usepackage{tcolorbox}

\usepackage{tcolorbox}    
\tcbuselibrary{breakable}  
\tcbuselibrary{skins}      

\usepackage{tabularx}
\usepackage{listings}

\title{\textsc{WisPaper}: Your AI Scholar Search Engine}

\author{
    Li Ju\textsuperscript{1,*},
    Jun Zhao\textsuperscript{2,*},
    Mingxu Chai\textsuperscript{2},  
    Ziyu Shen\textsuperscript{2},  
    Xiangyang Wang\textsuperscript{1},  
    Yage Geng\textsuperscript{1},  
    Chunchun Ma\textsuperscript{1},  
    Hao Peng\textsuperscript{1},  
    Guangbin Li\textsuperscript{1},  
    Tao Li\textsuperscript{1},  
    Chengyong Liao\textsuperscript{1},  
    Fu Wang\textsuperscript{1},  
    Xiaolong Wang\textsuperscript{1},  
    Junshen Chen\textsuperscript{1},  
    Rui Gong\textsuperscript{1},  
    Shijia Liang\textsuperscript{1},  
    Feiyan Li\textsuperscript{1},  
    Ming Zhang\textsuperscript{2},  
    Kexin Tan\textsuperscript{2}, 
    Junjie Ye\textsuperscript{2}, 
    Zhiheng Xi\textsuperscript{2},  
    Shihan Dou\textsuperscript{2},  
    Tao Gui\textsuperscript{2}, 
    Yuankai Ying\textsuperscript{1,2}, 
    Yang Shi\textsuperscript{1}, 
    Yue Zhang\textsuperscript{1},
    Qi Zhang\textsuperscript{1,2,$\dagger$}
}

\affiliation[1]{\mbox{WisPaper.ai}}
\affiliation[2]{\mbox{Fudan University}}

\abstract{
\begin{abstract}

We present \textsc{WisPaper}, an end-to-end agent system that transforms how researchers discover, organize, and track academic literature. The system addresses two fundamental challenges. (1)~\textit{Semantic search limitations}: existing academic search engines match keywords but cannot verify whether papers truly address complex research questions; and (2)~\textit{Workflow fragmentation}: researchers must manually stitch together separate tools for discovery, organization, and monitoring. \textsc{WisPaper} tackles these through three integrated modules. \textbf{Scholar Search} combines rapid keyword retrieval with \textit{Deep Search}, in which an agentic model, \textsc{WisModel}, validates candidate papers against user queries through structured reasoning. Discovered papers flow seamlessly into \textbf{Library} with one click, where systematic organization progressively builds a user profile that sharpens the recommendations of \textbf{AI Feeds}, which continuously surfaces relevant new publications and in turn guides subsequent exploration, closing the loop from discovery to long-term awareness. On TaxoBench, \textsc{WisPaper} achieves 22.26\% recall, surpassing the O3 baseline (20.92\%). Furthermore, \textsc{WisModel} attains 93.70\% validation accuracy, effectively mitigating retrieval hallucinations.
\end{abstract}
}

\correspondence{\email{qz@fudan.edu.cn}}
\checkdata[Website]{\url{https://wispaper.ai/}}

\begin{document}
\maketitle
\renewcommand{\thefootnote}{}
\footnotetext{$^*$Equal Contribution.\\$^\dagger$Corresponding authors.}
\renewcommand{\thefootnote}{\arabic{footnote}}


\vspace{-1.5em}

\section{Introduction}
The exponential growth of scientific publications presents unprecedented challenges for researchers attempting to stay current in their fields~\cite{article}. Efficiently identifying relevant work and systematically managing findings over time have become increasingly time-consuming and cognitively demanding~\cite{kingsley2011}. Two fundamental problems underlie this challenge.

The first is \textit{semantic search limitations}. Existing academic search engines such as Google Scholar\footnote{\url{https://scholar.google.com}} and Semantic Scholar\footnote{\url{https://www.semanticscholar.org/}} rely primarily on keyword-based retrieval, which struggles with conceptual queries such as ``\textit{papers exploring the relationship between in-context learning and inference-time scaling.}'' A researcher posing this question needs the system to understand what both concepts mean, find papers that engage with their intersection, and judge whether each 
paper truly contributes---not merely mentions the terms. Recent LLM-based approaches improve query reformulation~\cite{jagerman2023queryexpansionpromptinglarge, 10.1145/3539618.3591960, ma-etal-2023-query}, yet they still lack mechanisms for deep reasoning over paper content to verify actual relevance \cite{he-etal-2025-pasa}.

The second is \textit{workflow fragmentation}. A typical research workflow has three stages: discovering papers through search engines, organizing them in reference managers like Zotero\footnote{\url{https://www.zotero.org/}} or Mendeley\footnote{\url{https://www.mendeley.com/}}, and monitoring new publications via arXiv\footnote{\url{https://arxiv.org/}} or alert services. Each transition requires manual effort: downloading PDFs, copying metadata, tagging entries, checking for updates. Researchers spend significant time on information logistics rather than intellectual work \cite{Bullers2018}.

\vspace{-8pt}
\begin{figure}[h]  
    \begin{minipage}{0.6\columnwidth}
         We introduce \textsc{WisPaper}, an end-to-end scholarly agent system that addresses both challenges through three tightly integrated modules (Figure~\ref{fig:intro}). \textbf{Scholar Search} provides dual-mode retrieval: Quick Search for millisecond-level keyword matching, and Deep Search, an agent-powered mode where \textsc{WisModel}, a specialized LLM trained via reinforcement learning, decomposes queries into verification criteria and validates papers through structured reasoning. Discovered papers flow seamlessly into \textbf{Library} with one click, eliminating manual downloading and metadata entry. As papers accumulate, Library progressively builds a user profile that sharpens the recommendations of \textbf{AI Feeds}, which continuously surfaces relevant new publications and in turn guides subsequent exploration. Over time, each iteration of the cycle makes the next more efficient: collections grow richer, profiles become more precise, and recommendations align more closely with evolving research interests.
    \end{minipage}
    \hfill  
    \begin{minipage}{0.38\columnwidth}
        \centering
        \includegraphics[width=\linewidth]{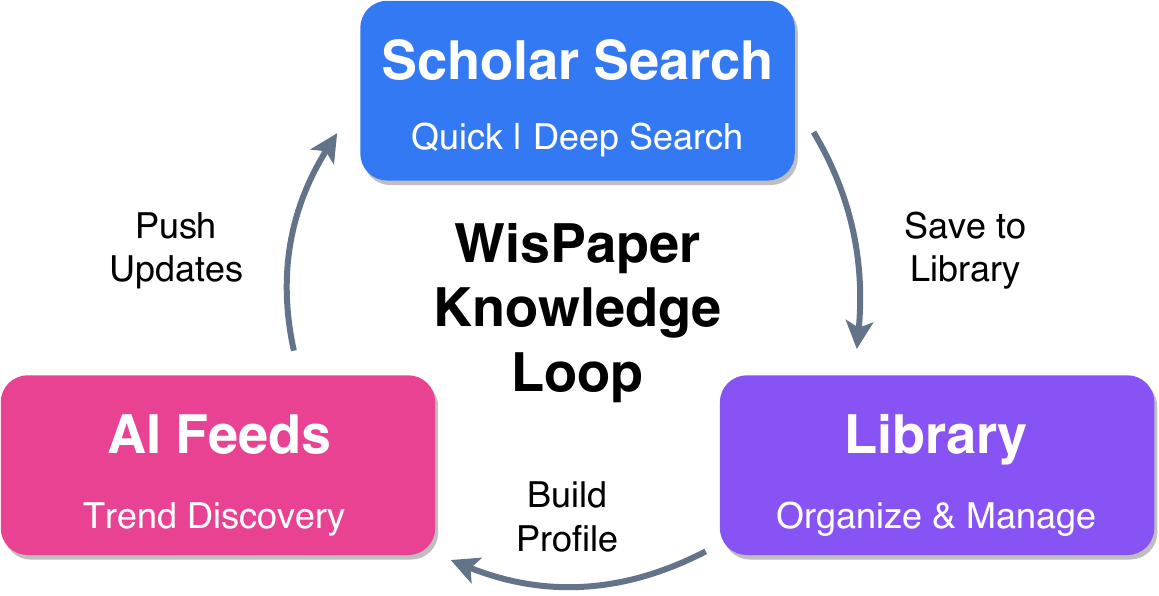}
        \caption{The \textsc{WisPaper} knowledge loop. Researchers discover papers via Scholar Search, organize them in Library, and receive continuous updates through AI Feeds. Modules share a unified paper database and exchange information to enhance each other.}
        \label{fig:intro}
    \end{minipage}
\end{figure}

Our main contributions are as follows:

(1) We present \textsc{WisPaper}, an end-to-end scholarly agent system in which search, organization, and monitoring form a closed knowledge loop.

(2) We develop \textsc{WisModel}, an agentic model trained via reinforcement learning that decomposes complex research queries into editable verification criteria and validates candidate papers through structured reasoning.

(3)  We demonstrate that \textsc{WisPaper} achieves 22.26\% recall on the TaxoBench benchmark and 93.70\% accuracy in criteria matching, surpassing existing state-of-the-art agents. 
\begin{figure*}[t]
    \centering
     \includegraphics[width=\linewidth]{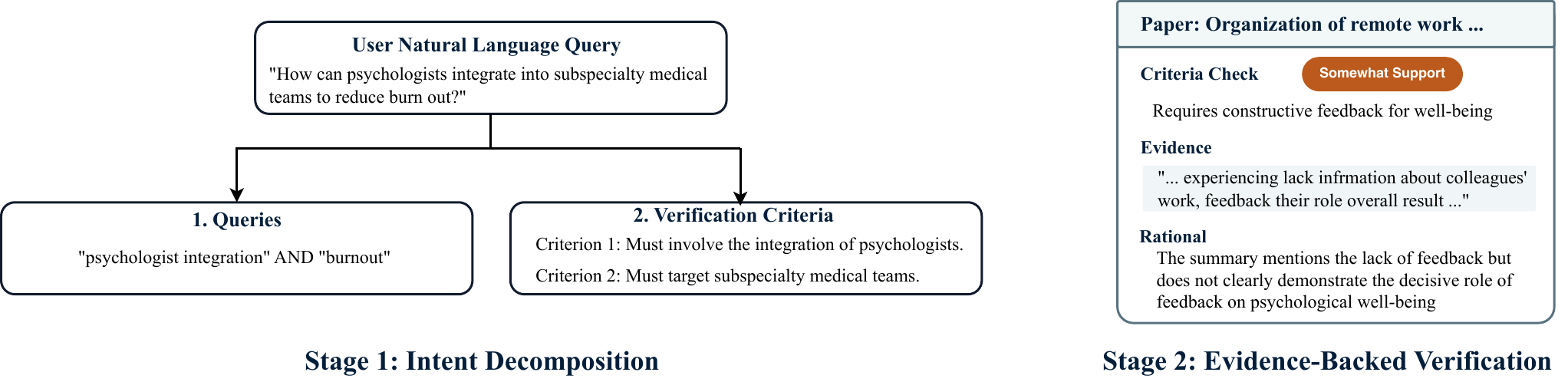}
    \caption{Deep Search workflow. In Stage~1, \textsc{WisModel} decomposes the user's natural language query into structured, executable queries and validation criteria. In Stage~2, the model validates candidate papers, strictly requiring direct quotes from the original text as evidence to eliminate hallucinations. Users can collaboratively refine criteria and re-execute the search.}
    \label{fig:deep_search_flow}
\end{figure*}

\section{The \textsc{WisPaper} Knowledge Loop}
\label{sec:overview}

\textsc{WisPaper} unifies literature discovery, organization, and monitoring into a single platform whose modules \textit{continuously exchange information to enhance each other}, forming a \textbf{knowledge loop}.

\paragraph{A Typical Session.}
A researcher's daily workflow illustrates the knowledge loop:
\textbf{(1)}~Open \textbf{AI Feeds} to review 3--5 papers recommended overnight (\S\ref{sec:ai_feeds}).
\textbf{(2)}~Add interesting papers to \textbf{Library} with one click; the system auto-extracts metadata (\S\ref{sec:library}).
\textbf{(3)}~While reading, encounter an unfamiliar technique; launch \textbf{Deep Search} directly from the PDF reader (\S\ref{sec:deep_search}).
\textbf{(4)}~Import newly found papers into Library, further refining the profile and improving tomorrow's AI Feeds.
Sections~\ref{sec:scholar_search}--\ref{sec:ai_feeds} detail each module.

\subsection{Scholar Search}
\label{sec:scholar_search}

Scholar Search offers two complementary modes for literature discovery.

\paragraph{Quick Search.} For well-defined information needs (e.g., looking up a specific paper by title or author), Quick Search provides millisecond-level keyword retrieval. Each result card displays title, authors, affiliations, venue, citation count, and an abstract snippet for rapid relevance judgments. Users can seamlessly import papers to Library with one click, copy formatted citations, and express relevance preferences (like/dislike) that feed into the user profile. 

\paragraph{Deep Search.}
\label{sec:deep_search}
Deep Search targets \textit{intent-heavy} queries that cannot be reliably answered by keyword matching (e.g., ``find recent AI4Science papers that \textit{evaluate} model generalization on real lab data''). It invokes \textsc{WisModel} (Section~\ref{sec:wismodel}) to run a two-stage pipeline that is \textbf{transparent} and \textbf{editable}:

\textit{Stage~1: Intent Decomposition.} \textsc{WisModel} first expands the user query into multiple distinct search \textit{Queries} to quickly retrieve a high-recall set of potentially relevant paper candidates. It then constructs an explicit \textit{verification checklist} that specifies what must hold for a paper to qualify (e.g., domain scope, method constraints). Both the search queries and the verification checklist are presented to the user as editable UI controls.

\textit{Stage~2: Evidence-Backed Verification.} For each retrieved candidate, \textsc{WisModel} produces a structured report that links each checklist item to \textit{concrete evidence} from the paper (e.g., title, abstract, and metadata) and an assessment label. Results are ranked by the aggregated checklist satisfaction, and the full per-criterion trace is exposed to support rapid inspection, thereby making each inclusion decision transparent and evidence-grounded.

\textit{Collaborative Refinement.}
Because the search queries and checklist are editable, users can revise retrieval terms and verification criteria and re-run verification without rephrasing the original query, progressively steering the search while keeping the system’s reasoning auditable.

\paragraph{Seeding the Knowledge Loop.} Discovered papers flow directly into Library: one-click paper import transfers full metadata to construct a personalized knowledge base, while simultaneously seeding the interest signals that refine the user profile within Library (\S\ref{sec:library}).

\subsection{Library}
\label{sec:library}
Library turns \textsc{WisPaper} from a search tool into a personal knowledge hub for papers. It streamlines paper ingestion and organization while continuously updating a user interest profile that drives the knowledge loop with AI Feeds.

\paragraph{Paper Import and Metadata Extraction.}
Papers enter Library through three pathways: direct PDF upload, one-click import from Scholar Search and AI Feeds. Upon ingestion, the system automatically extracts structured metadata (title, authors, affiliations, abstract, venue, date, DOI, and references) via PDF structure analysis, eliminating the manual entry that plagues traditional reference managers.

\paragraph{Organization and Export.} Library implements a hierarchical folder system with unlimited nesting (e.g., ``Survey $\rightarrow$ Retrieval $\rightarrow$ Dense Retrieval''), and supports both full-collection and subtree-constrained search, allowing users to locate papers by content rather than just by storage path. For seamless interoperability, it exports to standard bibliographic formats (BibTeX, RIS, EndNote) and offers synchronization plugins for Zotero and Mendeley.

\paragraph{User Interest Profiling.}
Library maintains an evolving interest profile that summarizes what the user is currently reading and collecting, which is subsequently consumed by AI Feeds (\S\ref{sec:ai_feeds}). The profile is updated from both \textit{explicit} signals (e.g., user-specified interests and curated folders) and \textit{implicit} behavioral signals (e.g., imported papers and how they are organized). Concretely, each newly added paper contributes topical/entity cues derived from its metadata (title, abstract, keywords, venue, authors), while the assigned folder path provides fine-grained, user-defined labels that disambiguate subtopics.

\paragraph{Driving the Knowledge Loop.} As papers accumulate and are organized, these import patterns and assigned folder hierarchies build a dynamic user interest profile that drives AI Feeds, making tomorrow's recommendations more precisely targeted.

\subsection{AI Feeds}
\label{sec:ai_feeds}

In fast-moving fields, researchers may spend 1–2 hours daily scanning journals and preprint servers. AI Feeds turns this passive monitoring into an active, personalized process through a dual-layer pipeline that dynamically adapts to the evolving user interests.

\paragraph{Dual-Layer Filtering.}
The first layer handles source selection: users subscribe to broad research areas (e.g., ArXiv categories such as \texttt{cs.CL}), and the system retrieves all new papers published since the last login, typically 50–100 per category per day. The second layer applies personalized filtering by matching these candidates against the user's current interest profile, narrowing daily papers to 3--5 highly relevant matches. Initially seeded by user-defined interests (e.g., ``vision-language models from Google/DeepMind''), this baseline profile is subsequently and continuously enriched by the Library module (\S\ref{sec:library}) as the user interacts with the system.

\paragraph{Completing the Knowledge Loop.}
AI Feeds completes the loop by channeling attention back into discovery: a recommended paper can be saved to Library with one click or can spark a new Scholar Search (\S\ref{sec:deep_search}), restarting the cycle with sharpened focus.


\section{\textsc{WisModel}: The Deep Search Agent}
\label{sec:wismodel}

Unlike standard conversational assistants that tolerate loose information retrieval, a scholarly agent faces a critical \textit{evidence-grounding challenge}: it must translate underspecified user intents into precise epistemological criteria and execute rigorous, hallucination-free validation over complex academic texts. 
Standard LLMs often fail at this, prone to ignoring nuanced constraints or fabricating supporting evidence. 
To overcome this, we formulate Deep Search as a structured reasoning task and train \textsc{WisModel} via a bespoke Reinforcement Learning (RL) pipeline to enforce both logical rigor and structural adherence.

\subsection{Task Formulation \& Initialization}
\label{sec:task_formulation}
We formalize the agent's reasoning process into two sequential objectives:

\paragraph{Objective 1: Intent Decomposition.} Given a raw user query $q$, the model generates a retrieval plan consisting of search queries $\mathcal{S}$ and a set of weighted verification criteria $\mathcal{C} = \{(c_i, w_i)\}_{i=1}^m$, where each $c_i$ represents a criteria and $w_i$ denotes its weight. Table~\ref{tab:example_query} provides a complete demonstration example of intent decomposition. The complete prompt template is provided in Table~\ref{tab:prompt_criteria_generation}.

\paragraph{Objective 2: Evidence-Backed Validation.} Given $q$, $\mathcal{C}$, and paper metadata $p$, the model acts as a rigorous reviewer, producing a structured verification report $y$. For each criterion $c_i$, the model must output a tuple $(a_i, e_i, r_i)$:
\begin{itemize}
    \item \textbf{Assessment} $a_i \in$ \{\textit{support}, \textit{somewhat support}, \textit{reject}, \textit{insufficient information}\}: A discrete logical verdict. The assessment taxonomy is shown in Table~\ref{tab:assessment_taxonomy}, with prompt templates in Tables~\ref{tab:prompt_validation} and \ref{tab:validation_output}.
    \item \textbf{Evidence Snippet} $e_i$: A verbatim quote extracted from $p$ that justifies $a_i$, strictly prohibiting hallucinated text.
    \item \textbf{Rationale} $r_i$: A concise natural language explanation connecting $e_i$ to $c_i$.
\end{itemize}

We initialize \textsc{WisModel} via Supervised Fine-Tuning (SFT) on an expert-annotated dataset. This stage bootstraps the model's ability to follow the complex JSON schema and understand the fundamental mapping between academic discourse and verification criteria.

\subsection{Reinforcement Learning via Multi-Dimensional Shaped Rewards}

While SFT establishes baseline capabilities, it often struggles to penalize subtle logical contradictions or hallucinated evidence---a common failure mode in academic verification. To address this, we further optimize the policy $\pi_\theta$ using Group Relative Policy Optimization (GRPO). GRPO samples $G$ output candidates $\{o_1, \ldots, o_G\}$ from the current policy $\pi_\theta$ and computes their respective rewards $\{r_1, \ldots, r_G\}$. We maximize the following objective:
\begin{equation}\label{grpo_loss}
    \mathcal{J}_{\text{GRPO}}\text{(}\theta\text{)} = \mathbf{E}_{x,\{o_i\}_{i=1}^G\sim\pi_{\theta_{\text{old}}}\text{(}\cdot|x\text{)}}\frac{1}{G}\vcenter{\hbox{\resizebox{.4cm}{!}{$\Sigma$}}}_{i=1}^G\frac{1}{|o_i|}\vcenter{\hbox{\resizebox{.4cm}{!}{$\Sigma$}}}_{t=1}^{|o_i|}\left[\frac{\pi_{\theta}\text{(}o_{i,t}|x,o_{i,<t}\text{)}}{\pi_{\theta_{\text{old}}}\text{(}o_{i,t}|x,o_{i,<t}\text{)}}A_{i,t}-\beta\mathbf{D}_{\text{KL}}\left(\pi_{\theta}||\pi_{\text{ref}}\right)\right],
\end{equation}
where $A_{i,t}=\text{(}r_i-\text{mean}\text{(}\mathbf{r}\text{))}\text{/std}\text{(}\mathbf{r}\text{)}$ is the advantage for each token. Instead of a sparse binary reward, we design a \textit{Multi-Dimensional Shaped Reward} function $\mathcal{R}(y)$ that decomposes the verification quality into four orthogonal dimensions:

\begin{enumerate}
    \item \textbf{Format Consistency ($r_{\text{fmt}}$):} Ensures deterministic system execution by rewarding strict adherence to the nested JSON schema and penalizing parsing errors.
    \item \textbf{Faithful Grounding ($r_{\text{ground}}$):} Specifically targets hallucination. It computes the longest common subsequence between the generated evidence $e_i$ and the source text $p$. If $e_i$ is not a substring of $p$, a severe penalty is applied.
    \item \textbf{Logical Entailment ($r_{\text{logic}}$):} Rewards the correctness of the discrete verdict $a_i$ against expert labels, heavily penalizing contradictions (e.g., predicting ``support'' when the paper explicitly ``rejects'').
    \item \textbf{Reasoning Alignment ($r_{\text{reason}}$):} Measures the semantic similarity of the generated rationale $r_i$ to expert explanations using embedding distance, ensuring the reasoning path is sound.
\end{enumerate}

\paragraph{Dynamic Curriculum Optimization.} 
Directly optimizing all rewards can lead to instability. We explicitly adopt a curriculum strategy by introducing dynamic time-dependent weights $\lambda_t$ for each reward component. The total reward at training step $t$ is defined as:
\begin{equation}
    \mathcal{R}_t\text{(}y\text{)} = \lambda_t^{\text{fmt}} r_{\text{fmt}} \text{+} \lambda_t^{\text{ground}} r_{\text{ground}} \text{+} \lambda_t^{\text{sem}} \text{(}r_{\text{logic}} \text{+} r_{\text{reason}}\text{)}
\end{equation}
Early in training, we prioritize $\lambda^{\text{fmt}}$ to enforce structural validity. As the KL divergence stabilizes, we decay $\lambda^{\text{fmt}}$ and increase $\lambda^{\text{ground}}$ and $\lambda^{\text{sem}}$, shifting the optimization focus from syntax to deep semantic reasoning. GRPO then updates the policy by favoring higher-reward trajectories relative to the group mean, effectively aligning the model with the rigorous standards of scientific inquiry.
\section{Evaluation}
We evaluate the two core capabilities underlying Deep Search: (1)~Intent Decomposition, (2)~Evidence-Backed Validation (\S\ref{sec:task_formulation}), and additionally (3)~end-to-end paper recall on the TaxoBench benchmark~\cite{zhang2026deepresearchagentsretrieve}, which measures whether the full Deep Search pipeline can retrieve the key papers that domain experts deem essential for a given research topic.

    \subsection{Evaluation Setup}

    \textbf{Data Collection and Annotation:} We recruited domain-expert PhD students to annotate real-world queries spanning 10 academic disciplines. Our dataset comprises 2,777 authentic queries in both Chinese and English, covering diverse research areas including computer science, biology, physics, economics, and social sciences. For each query, annotators provided:
    \begin{itemize}
        \item \textbf{Search queries:} Optimized keyword combinations for academic database retrieval.
        \item \textbf{Criteria:} Specific conditions to determine whether a paper falls within the target scope.
    \end{itemize}

\begin{figure*}[t]
    \centering
        \includegraphics[width=\linewidth]{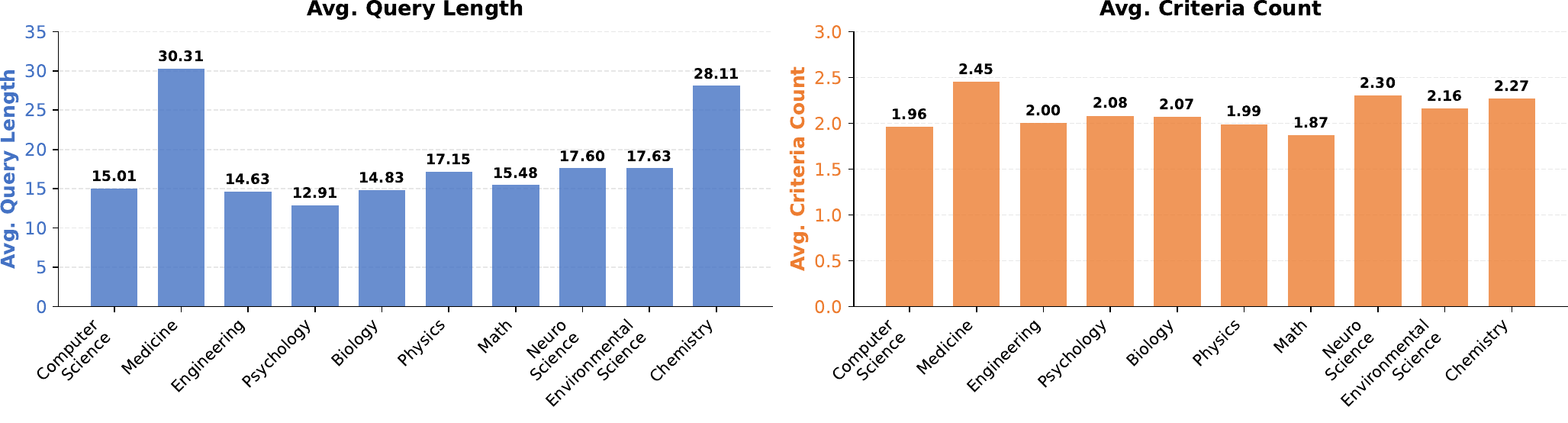}
        \caption{Distribution statistics of the evaluation dataset across 10 academic disciplines. The left panel shows the average query length in tokens for each discipline. The right panel displays the average number of criteria per query. These statistics indicate the complexity of search requirements across different fields.}
        \label{fig:domain_stat}
    \end{figure*}

    The distribution statistics are shown in Figure \ref{fig:domain_stat}, and representative query examples with their corresponding criteria are presented in Table \ref{tab:case}.

\textbf{Example Query-Criteria Pairs.} Table \ref{tab:case} illustrates the diversity of search intents across disciplines. For instance, a Computer Science query asks ``\textit{Are there any papers that investigate the relationship between in-context learning and inference-time scaling?}'', with criteria requiring papers to investigate both in-context learning methods and inference-time scaling methods. A Medicine query seeks ``\textit{Find the ELN guideline that recommends ruxolitinib for patients with myelofibrosis}'', with criteria specifying both guideline type (ELN) and treatment recommendation content. These examples demonstrate how WisModel must decompose complex queries into verifiable, independent criteria.

\begin{table*}[t]
    \footnotesize 
    \begin{tabularx}{\linewidth}{>{\centering\arraybackslash}p{0.12\linewidth} 
                                  >{\raggedright\arraybackslash}X 
                                  >{\raggedright\arraybackslash}X} 
    \toprule 
    \textbf{Domain} & \textbf{Query Example}& \textbf{Criteria Example}\\
    \midrule
    \rowcolor{gray!12}Computer Science & 
    Are there any papers that investigate the relationship between in-context learning and inference-time scaling? (e.g., 20 few-shot examples in prompt vs. self-training). & 
    1. The paper investigates in-context learning methods or their performance.\par
    2. The paper investigates inference-time scaling methods or their performance.\\
    
    Medicine & 
    Find the ELN guideline that recommends ruxolitinib for patients with myelofibrosis. &
    1. The paper is an ELN (European LeukemiaNet) guideline or consensus statement.\par
    2. The paper recommends or includes ruxolitinib as a treatment option for patients with myelofibrosis.\\
    
    \rowcolor{gray!12}Engineering & 
    DC arc fault detection method. & 
    1. The paper proposes or evaluates a method for detecting direct current (DC) arc faults.\\
    
    Psychology & 
    Find related papers about adults use physical experiences to understand and reason about spatial concepts. & 
    1. The paper investigates spatial reasoning or understanding in adults.\par
    2. The study uses physical experience or interaction as a method to explore spatial concepts.\\
    
    \rowcolor{gray!12}Biology & 
    Find papers on using Raman for DNA Sequencing. & 
    1. The paper discusses or applies Raman spectroscopy.\par
    2. The paper discusses or applies DNA sequencing.\\
    
    Physics & 
    Find papers that describe how to implement three position and three velocity into semi analytical models for galaxy/halo formation. &
    1. The paper describes or implements a method for incorporating three position and three velocity into semi-analytical models.\par
    2. The study focuses on galaxy or halo formation processes.\\
    
    \rowcolor{gray!12}Math & 
    I am searching for papers that treat the computation of p-values in mixture models. & 
    1. The paper proposes or discusses methods for computing or calculating p-values.\\
    
    Neuro Science & 
    Find papers about longtern Marajuan use and the brain & 
    1. The paper investigates the effects of long-term marijuana use.\par
    2. The study examines brain-related outcomes or changes.\\
    
    \rowcolor{gray!12}Environmental Science & 
    Are there any studies that have analyzed the seasonality of the Plant Area Index (PAI) to Leaf Area Index (LAI) ratio? & 
    1. The paper analyzes or discusses the seasonal variation or seasonality of the Plant Area Index (PAI).\par
    2. The paper analyzes or discusses the Leaf Area Index (LAI) ratio or its relationship with PAI.\\
    
    Chemistry & 
    Paper describing the contribution of chemical oscillations to the understanding of the dynamism of life phenomena. & 
    1. The paper discusses the role or contribution of chemical oscillations to the understanding of life phenomena.\par
    2. The paper addresses the dynamism or dynamic behavior of life phenomena.\\
    \bottomrule 
    \end{tabularx}
    \caption{Real-world queries and annotated criteria across 10 academic disciplines. Examples demonstrate the variety of search intents and the multiple criteria users specify when seeking relevant academic papers in different domains.}
    \label{tab:case}
\end{table*}

\subsection{Query Understanding and Criteria Generation}
\textbf{Task Description.} This experiment evaluates whether WisModel's generated outputs (search queries and criteria following the format in Appendix Table \ref{tab:prompt_criteria_generation}) align with human expert annotations. This corresponds to the first stage of Deep Search's agent workflow (Section \ref{sec:deep_search}).

\textbf{Evaluation Metrics:} We employ multiple complementary metrics:
\begin{itemize}
    \item \textbf{Semantic Similarity:} Cosine similarity using OpenAI's \texttt{text-embedding-3-large embeddings}\footnote{\url{https://platform.openai.com/docs/guides/embeddings}}, measuring semantic alignment between generated and reference texts.
    \item \textbf{ROUGE scores:} ROUGE-1, ROUGE-2, and ROUGE-L for n-gram overlap evaluation.
    \item \textbf{BLEU:} Precision-focused metric for exact phrase matching.
    \item \textbf{Length Ratio:} Percentage comparing generated text length to reference length (optimal: 100\%)
\end{itemize}
    
\textbf{Baseline Models:} We compare WisModel against state-of-the-art language models:
\begin{itemize}
    \item \textbf{GPT-5.1:} OpenAI's latest flagship model with adaptive reasoning mechanisms.
    \item \textbf{GPT-4o:} OpenAI's widely-adopted commercial model with multimodal capabilities.
    \item \textbf{Qwen-Max:} Alibaba's high-performance model with trillion-scale MoE architecture.
    \item \textbf{GLM-4-Flash and GLM-4.6:} Zhipu AI's models, with GLM-4.6 (355B parameters) leading in coding tasks.
    \item \textbf{DeepSeek-V3.2-Exp:} Leading open-source alternative with strong reasoning performance.
\end{itemize}

\textbf{Results:} 
Table \ref{tab:exp1} presents the comprehensive evaluation results. Key findings include:

(1) WisModel establishes new state-of-the-art: WisModel achieves dominant performance across all metrics, with Semantic Similarity of 94.8\%, ROUGE-L of 67.7\%, and BLEU of 39.8\%. This represents substantial improvements of 4.8\%, 15.1\%, and 18.3\% respectively over the second-best model.

(2) DeepSeek-V3.2-Exp emerges as the breakthrough open-source model: DeepSeek-V3.2-Exp demonstrates remarkable performance, achieving 90.2\% semantic similarity—surpassing several commercial models. This marks a significant milestone where open-source models match or exceed proprietary alternatives in specialized academic tasks.

(3) GPT-4o maintains reliability: GPT-4o delivers consistent performance (91.3\% semantic similarity, 52.6\% ROUGE-L, 21.5\% BLEU), making it a dependable choice for production deployments.

(4) Model-specific trade-offs emerge: While GPT-5.1 achieves strong semantic understanding (87.0\%), it struggles with exact phrase reproduction (BLEU 13.2\%). This suggests that newer models do not universally improve performance across all evaluation dimensions.

\begin{table}[t]
\centering
\begin{tabular}{lcccccc}
\toprule
\textbf{Model} & \textbf{\makecell[c]{Semantic\\Similarity}} & \textbf{ROUGE-1} & \textbf{ROUGE-2} & \textbf{ROUGE-L} & \textbf{BLEU} & \textbf{Length Ratio} \\
\midrule
Qwen-Max & 78.1 & 43.2 & 23.1 & 35.8 & 11.8 & 168.9 \\
GPT-4o & 91.3 & 64.0 & 39.4 & 52.6 & 21.5 & 142.2 \\
GPT-5 & 87.0 & 53.8 & 27.6 & 41.8 & 13.2 & 163.3 \\
GLM-4-Flash & 82.2 & 50.0 & 25.8 & 42.1 & 9.9 & 167.1 \\
GLM-4.6 & 84.8 & 55.5 & 30.2 & 44.5 & 14.4 & 168.1 \\
DeepSeek-V3.2-Exp & 90.2 & 59.3 & 32.4 & 48.0 & 14.4 & 153.5 \\
WisModel & \textbf{94.8} & \textbf{74.9} & \textbf{54.4} & \textbf{67.7} & \textbf{39.8} & \textbf{98.2} \\
\bottomrule
\end{tabular}
\caption{Model performance on query understanding and criteria generation task. Evaluation metrics include semantic similarity, ROUGE scores, BLEU, and length ratio. WisModel outperforms all baseline models across all dimensions.}
\label{tab:exp1}
\end{table}

\subsection{Paper-Criteria Matching}
\label{sec:exp2}
\textbf{Task Description:} This experiment evaluates WisModel's ability to accurately determine whether papers satisfy specified criteria, corresponding to the second stage of Deep Search (Section \ref{sec:deep_search}). Given the 2,777 queries and their 5,879 human-annotated criteria from Section 5.2, we evaluate matching accuracy across the four assessment categories defined in our taxonomy (Appendix Table \ref{tab:assessment_taxonomy}). Domain-expert PhD students labeled each paper-criteria pair using these categories.  The complete prompt structure for this task is provided in Appendix Tables \ref{tab:prompt_validation} and \ref{tab:validation_output}.

\textbf{Results:} Table \ref{tab:exp2} presents the matching accuracy for each category. Key findings include:

(1) WisModel demonstrates overwhelming superiority: WisModel achieves 93.70\% overall accuracy compared to the next best model at 73.23\%, with strong performance across all four categories (90.64\% insufficient information, 94.54\% reject, 91.82\% somewhat support, 94.38\% support).

(2) Category-specific baseline performance varies dramatically:
\begin{itemize}
    \item Support Recognition: Gemini-3-Pro excels at identifying fully supportive papers (91.1\%).
    \item Reject Detection: Qwen3-Max achieves the highest baseline rejection accuracy (72.0\%).
    \item Insufficient Information: Both Gemini-3-Pro and GPT-5.1 exceed 64\% accuracy.
    \item Somewhat Support - A Shared Weakness: Baseline models struggle significantly, with Gemini-3-Pro achieving only 15.9\%—a dramatic drop from its 91.1\% support accuracy.
\end{itemize}

(3) The ``partial support'' challenge: The poor performance on ``somewhat support'' reveals a critical limitation of existing models—they lack the nuanced judgment required for gray-area cases where papers partially address criteria. This represents a key capability that distinguishes WisModel from current alternatives.

\begin{table}[t]
\centering
\begin{tabular}{cccccc}
\toprule
\multirow{2}{*}{\textbf{Model}}&\multicolumn{4}{c}{\textbf{Criteria Categories}}& \multirow{2}{*}{\textbf{\makecell[c]{Overall\\Accuracy}}}\\
\cline{2-5}
& insufficient information & reject & somewhat support & support & \\
\midrule
GPT-5.1& 64.30 & 63.10 & 31.40 & 85.40 & 70.81 \\
Claude-Sonnet-4.5& 46.00 & 66.50 & 33.30 & 87.00& 70.62\\
Qwen3-Max& 40.80 & 72.00 & 44.20 & 87.20 & 72.82\\
DeepSeek-V3.2& 57.90 & 49.20 & 45.00 & 87.00& 66.82\\
Gemini3-Pro& 67.40 & 66.80 & 15.90 & 91.10 &73.23 \\
WisModel& \textbf{92.64} & \textbf{94.34} & \textbf{93.62} & \textbf{94.20} & \textbf{93.70}\\
\bottomrule
\end{tabular}
\caption{Model accuracy on paper-criteria matching task. Performance is measured across four classification categories and overall accuracy. WisModel significantly outperforms all baseline models, achieving 93.70\% overall accuracy compared to the next best model at 73.23\%.}
\label{tab:exp2}
\end{table}

\textbf{Application Recommendations for Baselines:}
While WisModel dominates in academic paper retrieval scenarios, users without access may consider task-specific alternatives shown in Table \ref{tab:model_recommendation}. For identifying ``full support'' cases, Gemini-3-Pro achieves the highest match rate (91\%). For rejection detection, Qwen3-Max leads with 72\% accuracy. For balanced overall performance, Qwen3-Max shows no extreme weaknesses across categories.

\begin{table}[h]
  \centering
  \resizebox{\linewidth}{!}{
  \begin{tabular}{lll}
    \toprule
    \textbf{Application Scenario}          & \textbf{Recommended Model}       & \textbf{Rationale}                          \\
    \midrule
    Identifying ``full support''             & Gemini-3-Pro            & Highest support match rate (91\%)           \\
    Identifying ``rejection''                & Qwen3-Max               & Highest reject match rate (72\%)            \\
    Assessing information sufficiency      & Gemini-3-Pro / GPT-5.1           & Both exceed 64\% accuracy                   \\
    Detecting partial support              & DeepSeek-V3.2 / Qwen3-Max        & $\sim$45\% accuracy, though suboptimal      \\
    Balanced overall performance           & Qwen3-Max               & No extreme weaknesses across categories     \\
    \bottomrule
  \end{tabular}
  }
  \caption{Model Recommendation for Different Application Scenarios}
  \label{tab:model_recommendation}
\end{table}

\subsection{End-to-End Paper Recall}
\label{sec:exp3}

\vspace{-8pt}
\begin{figure}[h]  
    \begin{minipage}{0.6\columnwidth}
    The preceding experiments evaluate Deep Search's internal components in isolation. This experiment assesses the full pipeline end-to-end: given only a research topic as input, can \textsc{WisPaper} retrieve the core papers that domain experts consider essential? 
    
\vspace{0.8em}

    \textbf{Benchmark}: We adopt TaxoBench~\cite{zhang2026deepresearchagentsretrieve}, a benchmark designed to evaluate deep research agents on expert-level literature survey tasks. TaxoBench contains 72 high-quality, highly-cited surveys (average citations: 354.5) from 8 computer science subfields, including multimodal learning, reinforcement learning, alignment, and agents. From these surveys, annotators with CS doctoral degrees manually extracted hierarchical taxonomy trees and mapped 3,815 core papers to specific branches. The benchmark's \textit{Deep Research Mode} provides only the survey topic as input and measures 
    \end{minipage}
    \hfill  
    \begin{minipage}{0.38\columnwidth}
        \centering
        \includegraphics[width=\columnwidth]{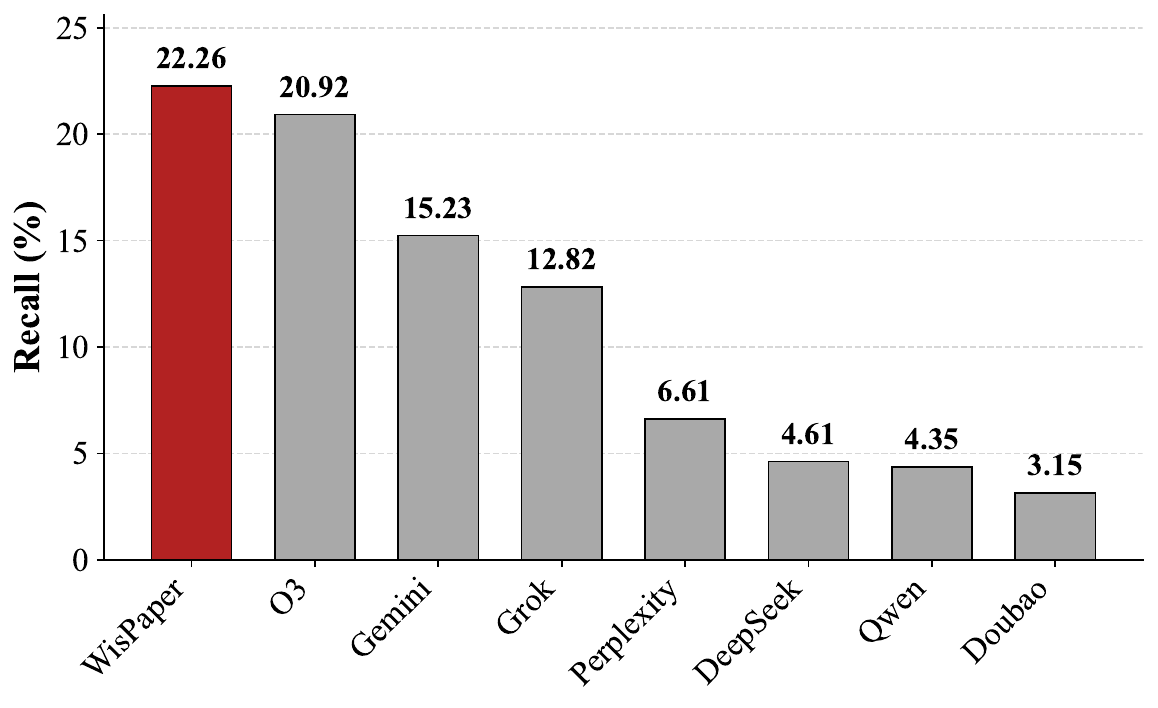}
        \caption{End-to-end paper recall (\%) on TaxoBench. \textsc{WisModel} achieves 22.26\%, surpassing the previous SOTA (O3, 20.92\%).}
        \label{fig:recall}
    \end{minipage}
\end{figure}
\vspace{-1.2em}
 the agent's ability to retrieve the 3,815 ground-truth papers---a challenging task on which the previous best system (O3) achieves only 20.92\% recall.

\textbf{Experimental Protocol}: We convert each survey title into a natural-language query (e.g., ``\textit{Find Papers about Evaluating Large Language Models in Code Generation Tasks}'' \(\rightarrow\) ``\textit{Find papers on Evaluating Large Language Models in Code Generation Tasks}'') and execute \textsc{WisPaper}'s Deep Search.
Retrieved papers classified as ``\texttt{Perfect}'' or ``\texttt{Partial}'' by \textsc{WisModel} are included in the result set.

\textbf{Results}: Figuer~\ref{fig:recall} presents the recall results. \textsc{WisPaper} surpasses the previous state of the art. \textsc{WisPaper} achieves 22.26\% Recall, exceeding the previous best result of 20.92\% set by O3 in TaxoBench's Deep Research Mode.
This is notable because O3 operates as a general-purpose deep research agent with unconstrained web access, whereas \textsc{WisPaper} retrieves from scholar indexed database, demonstrating the effectiveness of structured criteria-based validation.
\section{Related Work}
\subsection{Academic Retrieval and Discovery}
Academic search is a cornerstone of scientific research, with mainstream keyword-based systems such as Google Scholar\footnote{{\url{https://scholar.google.com/}}} and PubMed\footnote{\url{https://pubmed.ncbi.nlm.nih.gov/}} supporting global researchers. However, these tools often fall short in handling complex conceptual queries that demand semantic comprehension, as they rely primarily on surface-level text matching rather than deep insight into research intent \cite{https://doi.org/10.1002/jrsm.1378}. To address this limitation, recent studies have explored query reformulation and expansion techniques, yet these approaches still fail to break through the constraints of literal matching \cite{jagerman2023queryexpansionpromptinglarge, 10.1145/3539618.3591960, ma-etal-2023-query, 10.1145/3589335.3648298}. Subsequent advancements have introduced more sophisticated features: Semantic Scholar\footnote{\url{https://www.semanticscholar.org/}} integrated citation-based metrics and paper recommendation functions, while Connected Papers\footnote{\url{https://www.connectedpapers.com/}} enabled visual exploration of citation networks. These tools focus on relationship discovery but lack support for comprehensive literature management.

More recently, LLM-powered agents have brought new breakthroughs to academic search. LitSearch \cite{ajith-etal-2024-litsearch} and ResearchArena \cite{kang-xiong-2025-researcharena} have begun benchmarking LLM capabilities in this domain, verifying their potential to enhance retrieval performance. PaSa developed an autonomous agent that invokes search tools, parses papers, and navigates citation networks to address complex scholarly queries\cite{he-etal-2025-pasa}. PaSa outperforms baselines like Google Scholar and ChatGPT in recall metrics. Nevertheless, PaSa remains confined to the search phase and does not address post-retrieval needs such as literature organization and frontier tracking.

\subsection{LLM Agents for Research Assistance}
Recent efforts have shifted toward integrating discrete research tasks into cohesive workflows. Specialized systems like ChemCrow \cite{bran2023chemcrowaugmentinglargelanguagemodels} and Coscientist \cite{gottweis2025aicoscientist} automate chemistry experimentation, while virtual laboratory frameworks \cite{Swanson2024.11.11.623004} assist in wet-lab design. However, these tools are domain-specific and do not support cross-disciplinary literature management. End-to-end automation systems such as AI Scientist \cite{lu2024aiscientistfullyautomated} and ResearchAgent \cite{baek-etal-2025-researchagent} aim to cover ideation, experimentation, and paper writing. While innovative, their fully autonomous design limits user control and positions them as replacements rather than assistants for human researchers. Agent Laboratory \cite{schmidgall2025agentlaboratoryusingllm} expands the automated workflow to include literature review and report writing but prioritizes generating new research outputs over managing existing literature.
\section{Conclusion}
We presented \textsc{WisPaper}, a scholarly agent system designed to bridge the gap between information retrieval and knowledge management. By closing the loop between search, organization, and monitoring, \textsc{WisPaper} transforms passive literature consumption into an active, self-refining workflow. Central to our system is \textsc{WisPaper}, an agent explicitly trained to perform structured intent decomposition and rigorous evidence-backed verification. Our evaluation highlights the system's robustness: \textsc{WisPaper} achieves 93.70\% accuracy in paper validation, and the full pipeline attains a recall of 22.26\% on the TaxoBench benchmark, surpassing existing state-of-the-art agents. \textsc{WisPaper} is publicly deployed at https://wispaper.ai
, serving as a transparent and efficient tool for researchers across disciplines.

\section*{Acknowledgments}
We sincerely thank the following individuals for their valuable contributions to the data annotation process: Jiaqi Du (Nanjing University), Zixuan Wang (Fudan University), Xuetong Zhue (Tsinghua University), Tianjiao Xue (Tsinghua University), Shuyue Zhang (Sydney University), Feng Cui (Fudan University), Yonghao Liu (Zhejiang University), Yehao Zhang (Fudan University). Their dedicated efforts and high-quality work were essential to this research.

\clearpage

\bibliographystyle{plainnat}
\bibliography{main}
\clearpage
\section*{Appendix}

\begin{table*}[h]
\centering
\small
\begin{tabular}{p{0.18\linewidth}p{0.78\linewidth}}
\toprule
\textbf{Component} & \textbf{Content} \\
\midrule
\textbf{User Query} &
\textit{I am looking for articles that show how embedding psychologists into subspecialty medical teams reduce medical provider's level of burnout.''} \\ 
\hline 
\textbf{Generated} &  \textbf{Query 1:} embedded psychologist'' AND subspecialty medical team'' AND burnout \\
\textbf{Search Queries} & \textbf{Query 2:} psychologist integration'' AND ``medical team'' burnout \\
& \textbf{Query 3:} psychologist embedded medical team burnout \\
\hline
\textbf{Generated Criteria} &
\textbf{Criterion 1} (weight: 0.4, type: task) \\
& \quad \textit{Name:} Embedded psychologists \\
& \quad \textit{Description:} The paper examines the integration or embedding of psychologists. \\
\cline{2-2}
& \textbf{Criterion 2} (weight: 0.3, type: task) \\
& \quad \textit{Name:} Subspecialty medical teams \\
& \quad \textit{Description:} The paper is within subspecialty medical teams. \\
\cline{2-2}
& \textbf{Criterion 3} (weight: 0.3, type: task) \\
& \quad \textit{Name:} Impact on burnout reduction \\
& \quad \textit{Description:} The paper presents evidence or findings on the effect of embedded psychologists on reducing burnout among medical providers. \\
\bottomrule
\end{tabular}
\caption{End-to-end example of WisModel's query understanding and criteria generation (Objective 1). Given a natural language query about embedding psychologists in medical teams, WisModel performs query decomposition, generates three targeted Boolean search queries, and constructs three weighted validation criteria that operationalize the user's information need into verifiable checkpoints.}
\label{tab:example_query}
\end{table*}

\begin{table*}[t]
\centering
\small 
\begin{tabular}{p{0.18\linewidth}p{0.78\linewidth}} 
\toprule
\textbf{Component} & \textbf{Content} \\ \hline \textbf{User Query} &  \textit{Find a paper that argues remote workers who received limited feedback from supervisors experienced lower levels of motivation and job satisfaction and suggests that regular, constructive feedback is critical for maintaining engagement and enhancing employees' psychological well-being in remote work environments.''} \\
\hline
\textbf{Validation Criteria} &
\textbf{Criterion 1:} The paper examines the effects of limited feedback from supervisors on remote workers' motivation and job satisfaction. \\
\cline{2-2}
& \textbf{Criterion 2:} The paper argues that regular, constructive feedback is critical for maintaining engagement and enhancing employees' psychological well-being in remote work environments. \\
\hline
\textbf{Paper Metadata} &
\textbf{Title:} Organization of remote work in the context of digitalization \\
& \textbf{Authors:} Anna Sheveleva, Evgeny Rogov \\
& \textbf{Venue:} Developmental Psychology (Journal) \\
& \textbf{Year:} 2021 \\
& \textbf{Abstract (excerpt):} ...Individual-personal effects are connected with professional's personality transformation under influence environment digitalization: a change emotional side experiencing lack information about colleagues' work, feedback their role overall result...Organizational managerial associated implementation regulation, control assessment, staff motivation, forms employment, membership commitment, job satisfaction...'' \\ 
\hline 
\textbf{WisModel} &  \tikz\fill[green!60!black] (0,0) circle (0.8ex); \textbf{Criterion 1: Support} \\ 
\textbf{Assessment}& \quad \textit{Explanation:} The abstract discusses job satisfaction and motivation in the context of remote work, directly addressing the criterion. \\ & \quad \textit{Evidence:} Organizational managerial associated...staff motivation...job satisfaction, trust teams...'' \\
\cline{2-2}
& \tikz\draw[orange, line width=1.5pt] (0,0) circle (0.8ex); \textbf{Criterion 2: Somewhat Support} \\
& \quad \textit{Explanation:} The abstract mentions a lack of feedback (albeit not explicitly from supervisors), but does not explicitly argue that constructive feedback is critical for maintaining engagement or enhancing psychological well-being. \\
& \quad \textit{Evidence:} ``...experiencing lack information about colleagues' work, feedback their role overall result...'' \\
\hline
\textbf{Summary} &
The paper analyzes remote work organization in a digitalized context, exploring individual, social, and managerial effects such as lack of feedback and its potential impact on job satisfaction. However, it does not clearly argue that regular, constructive feedback is essential to maintain engagement or improve psychological well-being in remote work environments. \\
\bottomrule
\end{tabular}
\caption{Example of WisModel's paper-criteria validation (Objective 2) demonstrating nuanced assessment capabilities. Given a paper on remote work organization, WisModel correctly identifies \textit{support} for Criterion 1 (clear evidence present in abstract) and \textit{somewhat\_support} for Criterion 2 (indirect evidence only). This ability to distinguish between full and partial support represents a key strength—baseline models achieve only 15.9--45.0\% accuracy on somewhat\_support'' cases (Table~\ref{tab:exp2}), while WisModel reaches 91.82\%.} 
\label{tab:example_validation} 
\end{table*}

\begin{table}[h]
\centering
\small
\begin{tabular}{p{0.95\linewidth}}
\toprule
\textbf{System Prompt:} \\
\texttt{Your name is WisModel, an expert in academic search and literature screening. Your job is to:} \\
\texttt{1) Generate 2-4 Google Scholar search queries ("search\_queries").} \\
\texttt{2) Generate 1-4 executable, standalone screening criteria ("criteria"), each an independent rule.} \\

\textbf{User Prompt:} \\
\texttt{Current time: \{timestamp\}.} \\
\texttt{User query: \{user\_query\}} \\

\textbf{Expected Output Format (JSON):} \\
\texttt{\{} \\
\texttt{~~"search\_queries": [} \\
\texttt{~~~~"<Boolean search expression 1>",} \\
\texttt{~~~~"<Boolean search expression 2>",} \\
\texttt{~~~~...} \\
\texttt{~~],} \\
\texttt{~~"criteria": [} \\
\texttt{~~~~\{} \\
\texttt{~~~~~~"type": "<task|method|dataset|metric|etc.>",} \\
\texttt{~~~~~~"name": "<Short criterion name>",} \\
\texttt{~~~~~~"description": "<Detailed criterion description>",} \\
\texttt{~~~~~~"weight": <float between 0 and 1>} \\
\texttt{~~~~\},} \\
\texttt{~~~~...} \\
\texttt{~~]} \\
\texttt{\}} \\
\bottomrule
\end{tabular}
\caption{Prompt template for query decomposition and criteria generation (Objective 1).}
\label{tab:prompt_criteria_generation}
\end{table}

\begin{table}[h]
\centering
\small
\begin{tabular}{p{0.95\linewidth}}
\toprule
\textbf{System Prompt:} \\
\texttt{You are WisModel, a meticulous academic content auditor. Your task is to act as an academic expert and strictly follow a set of rules to verify if a given academic paper (paper\_info) aligns with a set of criteria derived from a user's query.} \\

\textbf{User Prompt:} \\
\texttt{Current time: \{timestamp\}} \\
\texttt{Original user query: \{user\_query\}} \\
\\
\texttt{**Validation criteria:**} \\
\texttt{<criteria>} \\
\texttt{~~<criterion\_1>\{criterion\_1\_description\}</criterion\_1>} \\
\texttt{~~<criterion\_2>\{criterion\_2\_description\}</criterion\_2>} \\
\texttt{~~...} \\
\texttt{</criteria>} \\
\\
\texttt{**Paper details for validation:**} \\
\texttt{<paper\_info>} \\
\texttt{~~<title>\{paper\_title\}</title>} \\
\texttt{~~<authors>\{paper\_authors\}</authors>} \\
\texttt{~~<affiliations>\{paper\_affiliations\}</affiliations>} \\
\texttt{~~<conference\_journal>\{venue\_name\}</conference\_journal>} \\
\texttt{~~<conference\_journal\_type>\{venue\_type\}</conference\_journal\_type>} \\
\texttt{~~<research\_field>\{research\_fields\}</research\_field>} \\
\texttt{~~<doi>\{paper\_doi\}</doi>} \\
\texttt{~~<publication\_date>\{publication\_date\}</publication\_date>} \\
\texttt{~~<abstract>\{paper\_abstract\}</abstract>} \\
\texttt{~~<citation\_count>\{citation\_count\}</citation\_count>} \\
\texttt{~~<source\_url>\{paper\_url\}</source\_url>} \\
\texttt{</paper\_info>} \\
\\
\texttt{**Your Task:**} \\
\texttt{Based on the rules provided in your instructions, you must perform a rigorous, step-by-step validation and generate a single JSON object as your response. Write all text fields (explanation, summary) in **English**.} \\
\bottomrule
\end{tabular}
\caption{Prompt template for paper-criteria alignment assessment (Objective 2).}
\label{tab:prompt_validation}
\end{table}

\begin{table}[h]
\centering
\small
\begin{tabular}{p{0.95\linewidth}}
\toprule
\textbf{Expected Output Format (JSON):} \\
\texttt{\{} \\
\texttt{~~"criteria\_assessment": [} \\
\texttt{~~~~\{} \\
\texttt{~~~~~~"criterion\_id": "<criterion\_id>",} \\
\texttt{~~~~~~"assessment": "<support|somewhat\_support|reject|insufficient\_information>",} \\
\texttt{~~~~~~"explanation": "<Natural language explanation for the assessment>",} \\
\texttt{~~~~~~"evidence": [} \\
\texttt{~~~~~~~~\{} \\
\texttt{~~~~~~~~~~"source": "<title|abstract|authors|etc.>",} \\
\texttt{~~~~~~~~~~"text": "<Extracted text supporting the assessment>"} \\
\texttt{~~~~~~~~\},} \\
\texttt{~~~~~~~~...} \\
\texttt{~~~~~~]} \\
\texttt{~~~~\},} \\
\texttt{~~~~...} \\
\texttt{~~],} \\
\texttt{~~"summary": "<Overall assessment summary explaining whether and how well the paper addresses the original query>"} \\
\texttt{\}} \\
\bottomrule
\end{tabular}
\caption{Expected output format for paper validation (Objective 2).}
\label{tab:validation_output}
\end{table}

\begin{table}[h]
\centering
\small
\begin{tabular}{p{0.12\linewidth}p{0.88\linewidth}}
\toprule
\textbf{Category} & \textbf{Definition and Examples} \\
\midrule
\textbf{Support} & 
The paper contains clear, direct evidence that confirms or fully satisfies the criterion. \\
& \textit{Example:} Criterion requires "model parameters exceeding 1 billion"; paper states "our model contains 1.2 billion parameters." \\
\hline
\textbf{Reject} & 
The paper contradicts the criterion or is fundamentally irrelevant. Two subcategories: \\
& (a) \textit{Explicit Contradiction:} Direct evidence negating the criterion (e.g., criterion requires "asynchronous training"; paper uses only "synchronous training"). \\
& (b) \textit{Foundational Irrelevance:} Domain/topic mismatch makes criterion inapplicable (e.g., criterion about "Chinese economic policy" applied to "Roman Empire agriculture"). \\
\hline
\textbf{Somewhat Support} & 
The paper partially relates to the criterion but provides incomplete, indirect, or limited evidence. \\
& \textit{Example:} Criterion requires testing on "ImageNet and COCO datasets"; paper only reports ImageNet results. \\
\hline
\textbf{Insufficient Information} & 
Available metadata (title, abstract, publication date) provides neither supporting nor contradicting evidence; relationship cannot be determined. \\
& \textit{Example:} Criterion asks about "experiments on physical robots"; abstract mentions neither platform details nor experimental setup. \\
\bottomrule
\end{tabular}
\caption{Assessment taxonomy and decision rules for paper validation.}
\label{tab:assessment_taxonomy}
\end{table}






\end{document}